\documentclass[12pt,a4paper]{scrartcl}

\usepackage{graphicx}
\usepackage{amsmath}
\usepackage{amsfonts}
\usepackage{amssymb}
\usepackage{amsthm}
\usepackage{color}
\usepackage[all]{xy}

\newcommand{\Spin}{{\rm Spin}}

\newcommand{\sll}{{\rm sl}}   
\newcommand{\SU}{{\rm SU}}
\newcommand{\DSU}{{\rm DSU}}

\newcommand{\SO}{{\rm SO}}

\newcommand{\su}{{\rm su}}
\newcommand{\Uq}{{\rm U}_q}
\newcommand{\dd}{{\mathrm d}}      
\newcommand{\Tr}{{\rm Tr}}

\newcommand{\R}{{\mathbb R}}   

\newcommand\sixj{6j--symbol}

\newcommand{\ph}[1]{\phantom{#1}}

\newcommand{\fig}[1]{Figure \ref{#1}}  

\theoremstyle{definition}
 
\begin{document}

\title{Holonomy observables in Ponzano-Regge type state sum models}

\author{John W. Barrett\\University of Nottingham\thanks{john.barrett@nottingham.ac.uk}\and
Frank Hellmann\\Albert Einstein Institute\thanks{hellmann@aei.mpg.de}}

\date{June 29th, 2011} 

\maketitle

\begin{abstract} We study observables on group elements in the Ponzano-Regge model. We show that these observables have a natural interpretation in terms of Feynman diagrams on a sphere and contrast them to the well studied observables on the spin labels. We elucidate this interpretation by showing how they arise from the no-gravity limit of the Turaev-Viro model and Chern-Simons theory.  
\end{abstract}

\section{Introduction}

The Ponzano-Regge model \cite{ponzanoregge} is a model for 3-dimensional quantum gravity without matter. The partition function is formulated as a state sum model, with a sum over labellings of a triangulated 3-manifold, and is independent of the triangulation chosen.

An important generalisation of this idea occurred when Witten proposed to construct 3-dimensional quantum gravity as a Chern-Simons type functional integral \cite{Witten1988a}. He developed both a version with a cosmological constant,
and a version without \cite{WittenTCA}. 

The model with a positive cosmological constant and Euclidean signature gives the same partition function as the state sum model constructed by Turaev and Viro \cite{Turaev:1992hq,Roberts}. 

This suggests that the Chern-Simons model without a cosmological constant should be the Ponzano-Regge model. Aspects of this have been confirmed \cite{Barrett:2008wh}; the Chern-Simons model giving a formula in terms of Ray-Singer torsion whereas the Ponzano-Regge formula gives a formula in terms of the equivalent Reidemeister torsion.
The model is a quantisation of the first-order form of 3-dimensional gravity, with action $\int tr(e \wedge F(\omega))$, where $e$ is an $\su(2)$-valued 1-form and $F$ is the curvature of the $\su(2)$ connection $\omega$. The functional integration is not subject to the condition that $\det(e)$ be positive, and thus $e$ behaves as a Lagrange multiplier. The functional integral thus reduces to an integration over the moduli space of flat $\su(2)$ connections.

Observables in these theories have been studied extensively. For the Turaev-Viro model a full account was given in \cite{Barrett:2004im}, for the Ponzano-Regge model in \cite{Barrett:2002vi,Freidel:2004vi,Barrett:2004gq,Freidel:2005bb,Barrett:2005vw}. These `edge observables' were always constructed by coupling spin labels on the edges in these state sum model. In this paper we consider instead observables of the $\SU(2)$ connection in the model. This leads to a different set of observables with a different physical interpretation, which is explained here.
To explore the interpretation fully, the new observables are considered as the $G\to0$ limit of observables for the Turaev-Viro model. This discussion is a development of the discussion in \cite{Freidel:2005bb}; we supply the observables which generate the Feynman diagram measure factors discussed there. The new observables are contrasted with the standard edge observables. The difference is that they are evaluations of spin networks for $\SO(4)$ and  $\DSU(2)$ respectively, quantum groups that are semi-dual in the sense of Majid-Schroers \cite{Majid:2008iz}.

\section{Group variables in the Ponzano-Regge model}
The Ponzano-Regge model \cite{ponzanoregge,Barrett:2008wh} on a triangulated 3-manifold $M$ can be expressed with spin variables $l_e\in\{0,\frac12,1,\ldots\}$ on each edge 
$e$  as
\begin{equation}\label{weightalt}
Z(M)=\sum_{l_e}\prod_{\text{interior edges}}(-1)^{2l_e}(2l_e+1) \prod_{\text{tetrahedra}}\left\{\begin{matrix}l_1&l_2&l_3\\l_4&l_5&l_6\end{matrix}\right\},
\end{equation}
the tetrahedral weight being the \sixj\ for the six spins $l_1\ldots l_6$ on its edges.
An alternative formula is obtained by writing the product of \sixj s as an integral over variables $g_f\in\SU(2)$ for each dual edge $f$, using the holonomy $h_e$ around each edge of the triangulation
\begin{equation}\label{firstorder}
Z(M)=\sum_l\prod_{\text{dual edges}}\int \dd g_f\; \prod_{\text{edges}}(2l_e+1)\Tr_{l_e}(h_e).
\end{equation}
Summing out the spin variables leads to a formulation entirely in terms of the group variables
\begin{equation}\label{groupvars} Z(M)=\prod_{\text{dual edges}}\int\dd g_f\;\prod_{\text{edges}}\delta(h_e).
\end{equation}
The delta functions on $\SU(2)$ (with support at the identity) force the $g$ variables to describe flat connections on $M$.

These expressions \eqref{weightalt},\eqref{firstorder} and \eqref{groupvars} all require regularizing; then they are equal for oriented compact manifolds in the circumstances when there are well-defined. They correspond to Lagrangian quantum gravity in the metric formulation, first-order formulation, and in a connection representation respectively. One needs to regularise \eqref{groupvars}, for example by systematically removing excess delta functions from the formula. For example in \cite{Freidel:2004vi}, the product over edges is restricted to a subset of edges excluding a maximal tree. For a full discussion of the divergences of this formula and their regularisation we point the reader to \cite{Barrett:2008wh} and \cite{Bonzom:2010ar,Bonzom:2010zh,Bonzom:2011br}. Such a regularisation will be assumed in the following.

Observables are usually inserted into the state sum model by including a function of the spin labels $j$ in the formula for the partition function $Z$ \cite{Barrett:2008wh,Freidel:2005bb,Freidel:2004vi}. These observables are called the `edge observables' in the following discussion, and are described more precisely in section \ref{edgesection}.

In this paper, the alternative possibility of using functions of the group variables is considered. This makes sense with either (\ref{firstorder}) or (\ref{groupvars}). As the observables don't depend on the spin labels one may as well sum these out. The formula with observables is thus
\begin{equation}\label{groupobs} Z(M,F)=\prod_{\text{dual edges}}\int\dd g_f \; F(g_1,g_2,\ldots)\; \prod_{\text{edges}}\delta(h_e).
\end{equation}
The observable is specified by the function $F$ of the variables $g_1,g_2,\ldots$ on the dual edges.

\subsection{Character observables on $S^3$}
The main features of the model with these observables are apparent in the special case where $M=S^3$ and $F$ is a product of character functions $\chi_j$ on $\SU(2)$ for some irreducible representation $j$ 
$$F=\chi_{j_1}(g_1)\chi_{j_2}(g_2)\chi_{j_3}(g_3)\ldots$$
The partition function is then a function of the spin labels $j_1,j_2,\ldots$ on some subset of the dual edges. These dual edges form a graph $\Gamma$ in $S^3$, and it is assumed that the edges of the graph are numbered with integers starting from 1. The graph together with the spin labels for its edges is denoted $\Gamma(j_1,j_2,\ldots)$.

The main mathematical result of this paper is the following identity.
\begin{equation}\label{RSN} Z(S^3, F)=(-1)^{\sum 2j_e}\left\langle \Gamma(j_1,j_2,\ldots) \right\rangle
\end{equation}
where $\left\langle \Gamma(j_1,j_2,\ldots) \right\rangle$ is the relativistic spin network evaluation of the labelled graph $\Gamma(j_1,j_2,\ldots)$ as defined in \cite{Barrett:1998dc}.

The proof of the formula follows from the fact that any flat connection on $S^3$ is pure gauge. Pick an arbitrary dual vertex $v_0$ of the triangulation as the origin. For each dual vertex $v$ define a new variable $u_v\in\SU(2)$ to be the product of group elements along some path that connects the dual vertex to the origin. This is well-defined since any two paths are homotopic via dual faces, and the delta function for a dual face ensures that going along different paths around the dual face gives the same group element.

The original group element on a dual edge $e$ with source and target vertex $s(e)$, $t(e)$ is recovered as 
$$g_e=u^{\ph{-1}}_{t(e)}u^{-1}_{s(e)}.$$ 
Using a maximal tree of dual edges, it is clear that the $u_v$ can take any values in $\SU(2)$, except that $u_{v_0}=1$, the identity element.

The integration measure after regularisation is
$$\prod_{\text{dual edges}}\dd g_f\;\prod_{\text{edges}}\delta(h_e)=\prod_{\text{dual vertices}}\dd u_v \,\delta(u_{v_0}). $$
Thus we have replaced the integration over flat connections with integration over the local gauge group. 

Dropping redundant delta functions, the overall partition function with the observables can now be written as
\begin{equation}\label{RSNeval} Z(S^3, F)= \prod_{\text{dual vertices of } \Gamma}\int\dd u_k \; \delta(u_{v_0})\; \prod_{\text{dual edges of } \Gamma}\chi_{j_{kl}}(u_k u^{-1}_l)\;
\end{equation}
Up to the minus signs, this is just the definition of relativistic spin network evaluation given in \cite{Barrett:1998dc}. This completes the proof of \eqref{RSN}.

\begin{figure}
\begin{center}
\includegraphics[scale=.5]{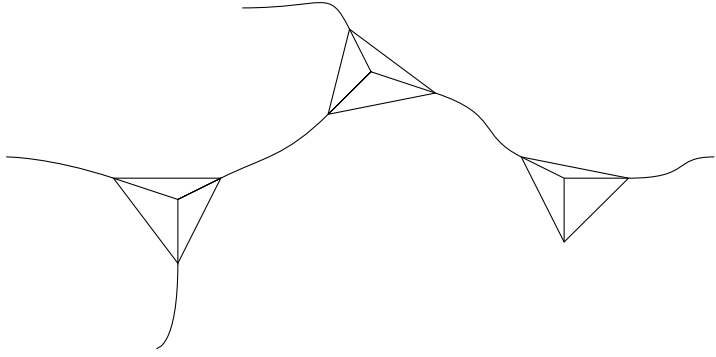}
\caption{Graph $\Gamma$ with expanded vertices.}\label{fig-graphwith6js}
\end{center}
\end{figure}

These observables can also be expressed as a modified state sum model. To see this, note that when integrating out the group elements on dual edges implementing the analogue of turning expression \eqref{firstorder} into \eqref{weightalt}, we no longer have three characters at the end of each dual edge which contract to form $6j$s, but instead, we get 4-valent intertwiners with one edge joined up.
\begin{equation}\label{6js}
\sum_{a_4} \int\dd g D^{l_1}_{a_1, b_1} D^{l_2}_{a_2, b_2} D^{l_3}_{a_3, b_3} D^{j}_{a_4, a_4} = \sum_{k, a_4} \iota_{a_1, a_2, a_3, a_4}^k \iota_{b_1, b_2, b_3, a_4}^k (-1)^{2k} (2k+1)
\end{equation}
where $k$ labels some basis of four-valent intertwiners. The state sum thus does not disconnect into a set of tetrahedral spin networks but into a set of tetrahedra joined by intertwiner labels and lines at the vertices. The result is that part of the state sum model becomes a spin network with the same topology as the graph $\Gamma$ in the observable, but with tetrahedral spin networks for vertices, see \fig{fig-graphwith6js}.

\subsection{General observables on $S^3$}

Above we assumed that we were dealing with observables that are
conjugation invariant and that thus can be expanded in characters. If
we chose $F^{gi}(g_1,g_2,\ldots)$ to be gauge invariant at the
vertices of the graph it can be expanded instead into spin networks.
Expressing the group observables at the edges through group elements
at the vertices again we see immediately that the observable simply is
the evaluation at the identity:

\begin{equation}
Z(S^3,F^{gi})= F^{gi}(1,1,\ldots)\, Z(S^3)
\end{equation}

A general function can only be decomposed into matrix elements of
representations. The matrix elements as observables on the group
elements constitute a generalization of relativistic spin networks.
The observable is in general a linear combination of observables of
the form
$$\prod_{\text{dual vertices}}\int\dd u_k \prod_{\text{dual edges}} \;
D^{j_{kl}}_{a_{kl} b_{lk}}(u_k u^{-1}_l),$$
with coefficients $O_{{kl}}^{a_{kl} b_{lk}}$ for each edge $kl$. Here
we think of the index $a_{kl}$ as living at the vertex $k$, facing the
edge $kl$.
The (almost) general observable on the group variables for fixed spins
can then be written in terms of these as
\begin{eqnarray}
Z(S^3,F)&=&\prod_{\text{dual vertices $i$}}\int\dd u_i \; \prod_{\text{dual
edges $(kl)$}} D^{j_{kl}}_{a_{kl} b_{lk}}(u_k u^{-1}_l)
O^{{kl}}_{a_{kl} b_{lk}} \nonumber\\
 &=&\prod_{\text{dual vertices $i$}}\int\dd u_i \prod_{\text{dual
edges $(kl)$}} \; D^{j_{kl}}_{a_{kl} c_{kl}}(u_k)
\epsilon_{c_{kl}d_{lk}} D^{j_{kl}}_{e_{lk} d_{lk}}(u_l)
\epsilon_{b_{lk}e_{lk}} O^{{kl}}_{a_{kl} b_{lk}},\label{gRSN}
\end{eqnarray}
where $\epsilon$ is the bilinear inner product on the $j_{kl}$ representation.

\begin{figure}
 \label{fig-relspinnetswithProps}
   \centering
   {   \def\svgwidth{0.7\columnwidth} 
   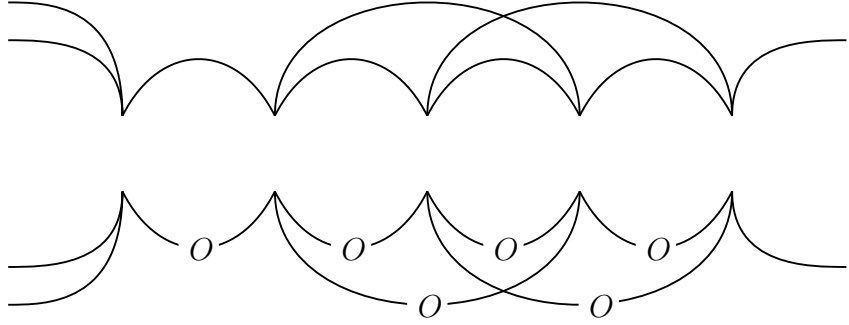}
\caption{Relativistic spin networks with operators $O$ along the edges.}
\end{figure}

We can now do the integration at each vertex and obtain the usual
projectors expressed as product of 3j symbols. Now note that from this
formula it is easy to see the the relativistic spin network evaluation
$\langle\cdot\rangle$ is up to signs equal to the square of the
ordinary spin network $\{\cdot \}$ for three-valent graphs. Setting
$O_{{kl}}^{a_{kl} b_{lk}} = \delta^{a_{kl} b_{lk}}$ and integrating
out the vertex variables $u_k$ locally by replacing them with
three-valent intertwiners gives exactly two sets of three-valent
intertwiners contracted according to the combinatorics of $\Gamma$.
Now restoring the $O_{(kl)}$, we obtain one normal spin network
valuation $\{\cdot \}$, and one network evaluation with operators
$O_{(kl)}$ used to contract intertwiners,
\begin{equation}\label{gRSN2} Z(S^3, \Gamma(j_1,j_2,\ldots, O))=
\left\{ \Gamma(j_1,j_2,\ldots) \right\}\left\{ \Gamma(j_1,j_2,\ldots,
O) \right\},
\end{equation}
see \fig{fig-relspinnetswithProps}. These observables are therefore
straightforward generalisations of relativistic spin networks.

\subsection{Generalising $S^3$}
Although initially formulated for a 3-manifold, the partition function (\ref{groupobs}) can be generalised by replacing the dual cell complex by an arbitrary cell complex $\mathcal K$. An example would be the dual cell complex of a 4-manifold, giving the Ooguri model \cite{Ooguri:1992eb}. The generalisation of (\ref{groupobs}) is
\begin{equation} \label{generalRSN} Z(M,F)=\prod_{\text{edges of $\mathcal K$}}\int\dd g_f \; F(g_1,g_2,\ldots)\; \prod_{\text{2-cells of $\mathcal K$}}\delta(h_e).
\end{equation}
The partition function is well-defined when a condition on the twisted cohomology is satisfied, as in \cite{Barrett:2008wh}. Cells of dimension higher than 2 play no role in the formula; however the partition function for a complex $\mathcal L$ containing higher-dimensional cells can be formulated by collapsing $\mathcal L$ to a 2-skeleton $\mathcal K$, and the regularisation of the Ponzano-Regge model can in fact be understood in this way. Thus formula \eqref{generalRSN} can be regarded as a generalisation of the definition of the relativistic spin network evaluation to an arbitrary manifold. The corresponding formula for $\Uq\sll(2)$ is studied in \cite{Barrett:2004im}

We have not studied the effect of arbitrary topology on the observables in a systematic way, but limit the discussion to a couple of examples. First, note that if the observable does not go around a non-contractible loop in $\mathcal K$ nothing in the analysis changes and the observable is the product of the $S^3$ observable with the graph $\Gamma$ and the evaluation of the partition function for $\mathcal K$ with no observable. To see this, simply run the change of integration argument above within this contractible region. Now the integration over vertices that touch the graph factorizes from the rest of the partition function. By introducing spurious integrations we can run the change of basis argument backwards while keeping the integrations touching the observable separate. This gives the product of the partition function and the integral on the right hand side of \eqref{RSNeval}. 

For the trivial case of a loop with one vertex and one 1-cell bounding a two-dimensional disk, the character observable \eqref{groupobs} is
$$\int\dd g\;\delta(g)\,\chi_j(g)=2j+1.$$

If $\mathcal K=S^1$, given by a loop with one vertex and one 1-cell with no disk filling it in, then the partition function with the character observable is
\begin{equation}\label{circle}\int \dd g\;\chi_{j}(g)=\delta_{j0}.\end{equation}
In essence, there is no propagation around the non-contractible loop. 

For $RP^2$, represented by a generator $g$ and a relation $g^2=1$, the corresponding partition function is
\begin{align}\int \dd g\; \delta(g^2)\chi_j(g)&=\frac18\left(\chi_j(1)+\chi_j(-1)\right)\\
&= \begin{cases}\frac14(2j+1) &\text{$2j$ even}\\0 &\text{$2j$ odd}\end{cases}\end{align}

These results will be interpreted in the next section.

\section{Particles}

\subsection{Particle on a Sphere}

The remaining task is to describe the physics of the new observables.

The first observation is that the new observables have the character of momentum observables. For example, if the graph $\Gamma$ has a 2-valent vertex, then the labels on either side are forced to be equal (else the partition function is zero). This is the conservation of momentum. For vertices of valence greater than 2, the restrictions on the values of the labels adjacent to a vertex are those compatible with the conservation of momentum where the labels are treated as the length of a momentum vector. 
Thus the labels can be considered as the absolute values of momenta, interpreted as (virtual) masses of particles.

One can see the interpretation of the partition function with the new observables directly from \eqref{RSNeval}. The $u$ variables can be considered as points on $S^3$. Thus one has the measure of a Feynman diagram amplitude. The character function on the group $SU(2)$ is the eigenfunction of the Laplacian on the homogeneous $S^3$ and thus the Feynman propagator (with a fixed virtual mass for the particle \cite{Barrett:2005vw}).  

This interpretation extends to the observables on manifolds other than $S^3$. For example for the case of $S^1$ considered above in \eqref{circle}
 interpreting $\chi_j(g)=\chi_j(g_t g_s^{-1})$ as a propagator we have the interpretation that the observable is given by the amplitude of propagation of a particle on the sphere from $g_s$ to the location $g_t$. The original observable is then recovered as the average amplitude of propagation over the whole of the space-time $S^3$, which vanishes except for $j = 0$.

\subsection{Edge observables}\label{edgesection}

A different set of observables was constructed in \cite{Freidel:2005bb,Barrett:2008wh}. These are the `edge observables'. This is defined by a graph $\Gamma$ consisting of edges of a triangulation, labelled by an angle $0\le \theta_e\le 2\pi$ associated to to each edge $e$. The amplitudes of these observables are then defined by inserting a factor 
$$K_\theta(l)=\frac{\sin \left(\frac{\theta} 2 (2l+1)\right)}{(2l+1)\sin\frac{\theta} 2}$$
 for each edge in the graph. This means the observable is
$$F(l_1,l_2,\ldots)=K_{\theta_1}(l_1)K_{\theta_2}(l_2)\ldots$$
and the Ponzano-Regge state sum \eqref{weightalt} with the edge observables is 
 \begin{equation}\label{edgeobs}
Z(F)=\sum_{l_e}\prod_{\text{interior edges}}(-1)^{2l_e}(2l_e+1) \prod_{\text{tetrahedra}}\left\{\begin{matrix}l_1&l_2&l_3\\l_4&l_5&l_6\end{matrix}\right\}\;F(l_1,l_2,\ldots).
\end{equation}
These observables are also momentum observables, conservation of momentum being respected at each vertex of the graph $\Gamma$. This is similar to the conservation of momentum for the new group observables, the principle difference being that in this case momentum space is curved. The partition function is zero unless there is, at each vertex,  a spherical polygon with side lengths the angles $\theta_e/2$ incident at the vertex. For a 2-valent vertex, the angles are required to be equal, expressing momentum conservation as before.

\subsection{Limits of the Turaev-Viro model}\label{TVsection}

The presence of two different momentum observables in the theory is at first puzzling. We will now elucidate their origin and differences by exhibiting both sets of observables as limits, as least heuristically, of the same observables for the Turaev-Viro model, in which the cosmological constant $\Lambda$ and the gravitation constant $G$ appear in complementary roles. Our thesis is that the edge observables are obtained from a $\Lambda\to0$ limit, and the new observables from a $G\to0$ limit of the Turaev-Viro model. Thus the edge observables pertain to quantum gravity without a cosmological constant whereas the new observables, at least locally, to quantum field theory on a three-sphere. 

\subsubsection{The limits}
The Turaev-Viro model is defined by a formula analogous to \eqref{weightalt}, with the \sixj\ and dimensions being replaced by their quantum deformations, which depend on an integer parameter $r$, also written as the deformation parameter $q=e^{i\pi/r}$. The quantum deformation of the dimension factor $(-1)^{2l}(2l+1)$ is
$$\dim_q l= (-1)^{2l}\frac{\sin\frac\pi r(2l+1)}{\sin\frac\pi r}.$$

The partition function is
\begin{equation}\label{TV}
Z_{TV}=N^{-v}\sum_{l_e=0}^{(r-2)/2}\prod_{\text{interior edges}}\dim_q l_e \prod_{\text{tetrahedra}}\left\{\begin{matrix}l_1&l_2&l_3\\l_4&l_5&l_6\end{matrix}\right\}_q.
\end{equation}
The main new features are the normalisation factor $N^{-v}$, with $v$ the number of vertices, and the dependence on the integer $r$, both explicitly in the limit of the sums, and via the deformation parameter $q=e^{i\pi/r}$. The model is finite and well-defined for any compact manifold, and is independent of the choice of triangulation.

Observables for the model are defined by multiplying the summand of the partition function by a function $F(l_1,l_2,\ldots)$ of some of the $l$s in the partition function\cite{Barrett:2002vi}. The $l$ variables on which $F$ depends are those lying on some subgraph $\Gamma$ of the edges of some triangulation. 

The relevant observables for this paper are the `momentum space' observables, where the function introduced is the product of a Fourier kernel 
$$\overline K_j(l)=(-1)^{2j}\frac{\sin\frac\pi r(2j+1)(2l+1)}{\sin\frac\pi r(2l+1)}$$
for each edge, so 
$$F=\overline K_{j_1}(l_1)\overline K_{j_2}(l_2)\ldots$$

As announced in \cite{Barrett:2002vi}, and proved in \cite{Barrett:2004im}, the partition function for $S^3$ with these observables is equal to a constant times the $q$ version of the Relativistic Spin Network invariant \cite{Barrett:1998dc,Crane:1997wt,Yokota199677} of the graph $\Gamma$ with its edges labelled with $j_1,j_2,\ldots$
\begin{equation}\label{qRSN}  Z_{TV}(S^3,\Gamma)=Z_{TV}(S^3)\,<\Gamma(j_1,j_2,...)>_{q}.\end{equation}

Heuristically the Turaev-Viro model, which is a quantisation of 3-dimensional GR with cosmological constant, can be seen to reduce to the Ponzano-Regge state sum when one takes the limit $r\to\infty$. However this limiting process is subtle and there are two different ways in which one can take a limit of the Turaev-Viro momentum space observables.

\begin{itemize}
\item[(i)] $$r \to\infty, \quad j_i  \text{ constant}$$
\item[(ii)]  $$r \to\infty, \quad \frac {2j_i+1}r\to\frac{\theta_i}{2\pi} \text{, a constant}$$
\end{itemize}
Limit (i) is the limit which gives the observables in this paper, as is clear from the limit
$$<\Gamma(j_1,j_2,...)>_{q}  \to <\Gamma(j_1,j_2,...)>,$$
which is obvious from the defining spin network formulae.

The second limit, (ii), is somewhat harder to treat rigorously. It is calculated explicitly for $\Gamma$ a trefoil knot in \cite{Barrett:2004gq} and aspects are generalised to any knot in \cite{Barrett:2008wh,Dubois} 
It seems a reasonable conjecture that when all the $\theta_e$ are sufficiently small, then a limit of the Turaev-Viro partition function with the Fourier kernel observables gives the Ponzano-Regge partition function with the edge observables. This is consistent with the fact that in this limit, and with $l$ fixed,
$$\frac{\overline K_j(l)}{\dim_q l}\to K_\theta(l).$$

\subsection{Physical parameters in the limits}
\subsubsection{The minimum mass}
To understand the physics of these limits we will introduce the dimensionful parameters into the theory. In the Turaev-Viro partition function without observables there is just one parameter, $r$. On the other hand quantum gravitational physics would seem to require three parameters, the gravitation constant $G$, Planck's constant $\hbar$ and the cosmological constant $\Lambda$. The constant $r$ can be written in terms of these three parameters but then it would seem that two of the parameters are redundant.

The resolution of this paradox is that the two additional parameters play a role when observables are introduced. In fact in three dimensions $1/G$ is a unit of mass and both $\hbar G$ and $1/\sqrt\Lambda$ are units of length. Therefore one can multiply the purely numerical measures of mass or length in Turaev-Viro observables by one of these units to get `physical' masses or lengths. The reason this is worthwhile is that in considering the scaling behaviour, it is more useful to consider the physical masses (for example) to be fixed and scale $G$, than it is to consider $G$ fixed and scale the numerical masses. This means that the physical mass converges to a definite value in a scaling limit (rather than, say, 0 or $\infty$).

In this paper, scales of lengths are not considered. Since the role of $\hbar$ is to relate the scale of masses with the scale of lengths, its scaling is irrelevant. It can be assumed to be a constant throughout. The two units of interest are therefore the two mass scales $1/G$ and $\hbar\sqrt\Lambda$.

In the Turaev-Viro model, the numerical values of the masses for the momentum observables range from the minimum $j+1/2=1/2$ to maximum $j+1/2=(r-1)/2$, the ratio between them being $r$, to leading order. The physical models for these masses\cite{Barrett:2002vi} are the zonal spherical functions on $S^3$, which for a sphere of radius $1/\sqrt\Lambda$ have  masses $m=\hbar\sqrt\Lambda(2j+1)$, using as definition for mass $m$ the eigenvalue equation $$\nabla^2\phi=\left(-\frac{m^2}{\hbar^2}+\text{const.}\right)\phi.$$
 Thus the numerical masses are multiplied by the physical unit $2\hbar\sqrt\Lambda$. This unit  is considered to be of cosmological origin, since the minimum mass corresponds to a particle with wavelength the circumference of $S^3$.

\subsubsection{The maximum mass}
 The model of a particle wavefunction on a classical geometry given in the previous section does not account for the maximum mass $\hbar\sqrt\Lambda(r-1)$ in the Turaev-Viro model. A completely different argument based on general relativity can be used to identify this maximum mass in terms of $G$.

The Einstein equation in three dimensions is written as
$$G_{\mu\nu}-\Lambda g_{\mu\nu}=8\pi G T_{\mu\nu},$$
the convention being that the constant $8\pi G$ is the same as in four dimensions.

The model for a particle is a conical defect in a spherical geometry. From the Einstein equation, the defect angle is given by $8\pi Gm$ \cite{DeserJackiwt'Hooft}. Since the defect angle is less than $2\pi$, the maximum mass is therefore just below $1/4G$. Hence
$$\frac1{4G}=\hbar\sqrt\Lambda\,r$$
This can be rearranged to give
\begin{equation}\label{rformula} r=\frac 1{4 \sqrt\Lambda G\hbar}.\end{equation}
 
Now it is possible to rewrite the limits using the physical constants. The physical mass on the $i$-th edge is defined as
$$m_i=(2j_i+1)\hbar\sqrt\Lambda=\theta_i/8\pi G,$$
and according to the above argument, $\theta_i$ is the defect angle of the corresponding geometry.

The two limits are
\begin{itemize}
\item[(i)] $m_i$, $j_i$, $\Lambda$ constant, $G\to 0$, $\theta_i\to 0$.
\item[(ii)]  $m_i$, $\theta_i$, $G$ constant, $\Lambda\to 0$, $j_i\to \infty$.
\end{itemize}

\subsection{The Functional Integral}
 A more precise relation with the physical constants can be determined using the   
 functional integral picture. The Turaev-Viro model can be written as a functional integral with action given by the difference of two Chern-Simons actions for $\SU(2)$ connections $A_a^+$ and $A_a^-$ \cite{Witten1988a},
$$ S=\frac k{4\pi}\int CS(A_a^+)-CS(A_a^-),$$
where a is an $\su(2)$ Lie-Algebra index, that is $[A_a^\pm, A_b^\pm] = \epsilon_{abc} A^{c \pm}$, and
$$CS(A)= \int \Tr (A \wedge dA + \frac23 A \wedge A \wedge A)$$

 Introducing the physical scales of a gravitation constant $G$, Planck constant $\hbar$ and cosmological constant $\Lambda$ by
$$k=\frac 1{4 \sqrt\Lambda G\hbar}$$
and changing variables to the usual fields of first order gravity, 
$$\omega_a=\frac12(A_a^++A_a^-)$$
$$e_a=\frac1{2\sqrt{\Lambda}}(A_a^+-A_a^-)$$
gives the familiar form
\begin{equation}\label{gravityaction}S=\frac 1{4\pi G\hbar} \int e_a\wedge d\omega^a + \epsilon^{abc} e_a \wedge \omega_b \wedge \omega_c + \frac{\Lambda}{3} \epsilon^{abc} e_a \wedge e_b \wedge e_c\\
=\frac{\pm1}{16\pi G\hbar}\int(R-2\Lambda)\dd V\end{equation}

Of course at this stage, two of the three scales are redundant, since the partition function without observables contains only one parameter, $r$. But the momentum observables considered here introduce a second parameter and the length observables (if we included them) would introduce a third parameter.

An observable for the Chern-Simons functional integral framework is given by a generalised Wilson loop that is supported on a graph. It is specified by a representation of the gauge group for each edge of the graph an an intertwining operator for each vertex. According to Witten \cite{WittenQFTJP}, the expectation value of this observable, in the case of $S^3$, is the corresponding quantum group evaluation for this data labelling a plane projection of the graph, using also the $R$-matrix for the quantum group at crossings. This evaluation is called a spin network evaluation for the quantum group.

The quantum group relativistic spin network evaluation $<\Gamma(j_1,j_2,...)>_{q}$, which appears in \eqref{qRSN}, is an example of a spin network evaluation \cite{Yetter}. The quantum group is $U_q\sll2\times U_q\sll2$ (with one factor using $q^{-1}$ in place of $q$), and the representations are $(j,j)$. The Wilson loop observable $W(j_1,j_2,\ldots)$ for the functional integral that generates this expectation value is therefore the function of the connection given by representing each edge as the parallel transport operator for $\SU(2)\times\SU(2)$ in the representation $(j,j)$, and each vertex by the canonical intertwining operator for the $\SU(2)$ relativistic spin networks.

The functional integral with the observable is
$$Z(W)=\int  [\dd e] [\dd \omega] e^{iS}W(j_1,j_2,\ldots).$$

The $G\to0$ limit is explained in a conceptual way using this formula. As the $j$ in the observable are fixed in this limit, it simply consists of taking the semi-classical limit in the functional integral. For this, the functional integral can be replaced by an integral over the space of classical fields given by the critical points of the effective action. The effective action for Chern-Simons is the same formula as the classical action, but with $k$ replaced by $r=k+2$ \cite{Alvarez-Gaumeetal}. This means that the physical constants are related by \eqref{rformula}, as before.

For the case of $S^3$, the critical points of the effective action are all gauge-equivalent to the trivial connection, and so, since $W(j_1,j_2,\ldots)$ is gauge invariant, the functional integral just amounts to evaluating $W$ on the trivial connection. This is just the alternative definition of the relativistic spin network for $\SU(2)$, and so leads to formula \eqref{RSN}.

Note that we can make direct contact to the formulation of the observables in the Ponzano-Regge context. To see this consider the observable given by inserting $\chi_j(g_e^+ {(g_e^-)}^{-1})$ into the Chern-Simons path integral for each edge of the graph, where $g_e^\pm$ is the parallel transport with respect to $A^\pm$ along $e$. As the path intgeral is gauge invariant this observable evaluates to the same as  $\chi_j(g_v g_e^+ g_{v'} {(g_e^-)}^{-1})$. By gauge averaging over the $g_v$ we obtain the observable $W(j_1,j_2,\ldots)$. On the other hand the critical points are simply the connections gauge equivalent to the trivial one, so the non gauge invariant form of the observable reduces to that we considered in the Ponzano-Regge model.

The $\Lambda\to0$ limit by contrast is not a semiclassical limit. Setting $\Lambda=0$ in the gravitational action \eqref{gravityaction} gives the Chern-Simons action for $ISU(2)$ \cite{WittenTCA}. This is an In\"on\"u-Wigner contraction of the gauge group, which suggests the corrresponding contraction of the representation matrix elements \cite{WignerI}. In such a contraction, the representation labels are scaled simultaneously with the group contraction so that the matrix elements of the representation converge. This is exactly the situation in the $\Lambda\to0$ limit; however studying this systematically would stray too far from the main aim of this paper.

\section{Discussion}

In this paper we discussed observables coupled to the group elements in the holonomy formulation of the Ponzano-Regge model. Specializing to observables that are products of characters we showed that we recover the relativistic spin network evaluation. This also shows very simply that these have an interpretation as the evaluation of momentum labelled Feynman diagrams on the sphere.

The previously considered observables coupled to the spin labels also have the character of momentum observables, we elucidate the presence of two types of momentum observables by showing that they arise as two different limits of the momentum observables in the Turaev-Viro model. To understand the physics of this limit we reintroduced the dimensionful quantities. This demonstrates that the observables introduced in this paper, and the Ponzano-Regge model on the whole, can be understood as the semiclassical limit of the Turaev-Viro model.

To further see how this happens we consider the Chern-Simons path integral formulation of the Turaev-Viro model. The analogue of the holonomy observables can be seen to be given by introducing characters depending on the product of holonomies $g^+ {(g^-)}^{-1}$. This has a natural interpretation in terms of the geometry of a sphere as it parametrizes the coset space of $\Spin(4) \cong \SU(2)\times \SU(2)$ with respect to the diagonal subgroup $\SU(2)_d$. This coset space is just the homogeneous 3-sphere with $\Spin(4)$ as its the global group of symmetries, our scalar observables do not see the local rotational symmetry $\SU(2)_d$ but only the translational part.

This suggests that it might be possible to interpret these observables on general 3-manifolds as particles propagating in a locally flat Cartan geometry modelled on $\SU(2) \times \SU(2)/\SU(2)_d \cong S^3$ on that manifold. The overall amplitude is then obtained by averaging over the moduli space of flat Cartan geometries.

One can also consider doing a further limit that takes us from the model based on $\SU(2)$ to the abelian model based on $\R^3$. These should be the $G\to0$ limit for the edge observables which can be considered as a semiclassical limit. This was discussed as a commutative limit of the effective field theory in \cite{Freidel:2005bb}, and the $\Lambda\to0$ limit of the group observables, which should give the same physics again in a dual picture. As conjectured above this should be achieved by an In\"on\"u-Wigner contraction of the group $\SU(2)$.

Note that in this work we find some of the dualities found by Majid and Schroers in \cite{Majid:2008iz} at the level of the particles propagating on (non-commutative) space times at the level of the state sums. In particular the dual limits of the Turaev-Viro observables lead to spin network evaluations that are semi-dual in the sense of \cite{Majid:2008iz}.

\bibliography{gvbib}
\bibliographystyle{hep}

\end{document}